\documentclass[aps,pra,10pt,showpacs,twocolumn,longbibliography,superscriptaddress]{revtex4}
\usepackage{threeparttable}
\usepackage{indentfirst}
\usepackage{amsmath}
\usepackage{lipsum}
\usepackage{multirow}
\usepackage{booktabs}
\usepackage{graphicx,color}
\usepackage{mathrsfs}
\usepackage{graphicx}
\usepackage{subfigure}
\usepackage{amsmath}
\usepackage{mathtools}
\usepackage{braket}
\usepackage{color}

\newcommand{\ba}{\begin{eqnarray}}
\newcommand{\ea}{\end{eqnarray}}

\newcommand{\be}{\begin{equation}}
\newcommand{\ee}{\end{equation}}

\definecolor{pink}{rgb}{1,0.18,1.0}

\def\2dm{{2D Materials}}

\begin{document}

\title{Strong and nearly 100$\%$ spin-polarized second-harmonic generation from ferrimagnet Mn$_{2}$RuGa}

\author{Y. Q. Liu}
\affiliation{School of Materials and Energy, Lanzhou University, Lanzhou 730000, China}

\author{M. S. Si$^{*}$}
\affiliation{School of Materials and Energy, Lanzhou University, Lanzhou 730000, China}

\author{G. P. Zhang$^{\dagger}$}
\affiliation{Department of Physics, Indiana State University, Terre Haute, IN 47809, USA}

\date{\today}
\begin{abstract}

Second-harmonic generation (SHG) has emerged as a promising tool
for detecting electronic and magnetic structures in noncentrosymmetric materials,
but 100$\%$ spin-polarized SHG has not been reported.
In this work, we demonstrate nearly 100$\%$ spin-polarized SHG from half-metallic ferrimagnet Mn$_{2}$RuGa.
A band gap in the spin-down channel suppresses SHG,
so the spin-up channel contributes nearly all the signal, as large as 3614 pm/V about 10 times larger than that of GaAs.
In the spin-up channel, $\chi_{xyz}^{(2)}$ is dominated by the large intraband current in three highly dispersed bands near the Fermi level.
With the spin-orbit coupling (SOC), the reduced magnetic point group allows additional SHG components, where
the interband contribution is enhanced.
Our finding is important as it predicts a large and complete spin-polarized SHG in a all-optical spin switching ferrimagnet.
This opens the door for future applications.

\end{abstract}

\maketitle

\section{Introduction}
The interaction between an intense optical field and a material is always fascinating.
This gave birth to nonlinear optics \cite{SHG1,SHG2}.
Second-harmonic generation (SHG), a special case of sum frequency generation, has received enormous attention worldwide.
SHG only exists in noncentrosymmetric materials with broken inversion symmetry \textit{I} \cite{noncentro1},
while it is absent in centrosymmetric systems.
In general, impurities and surfaces introduced in a material can break \textit{I}.
For instance, \textit{I} of the NV center is broken by introducing nitrogen-vacancies in diamond \cite{NV1,NV2,NV3,NV4}.
Moreover, a few layers of crystal, created by mechanical exfoliation, exhibit different symmetry properties.
Odd layers of MoS$_{2}$ and h-BN belong to the noncentrosymmetric space group,
different from their bulk, can also generate SHG \cite{MoS21,MoS22,MoS23}.
So far, most of the materials studied are nonmagnetic.
For magnetic materials, magnetic order can break time reversal symmetry \textit{T}.
The sizable SHG appears in the antiferromagnetic (AFM) CrI$_{3}$ and
the even septuple layers of MnBi$_{2}$Te$_{4}$ \cite{CrI1,CrI2,MnBiTe1},
where the AFM ordering breaks \textit{I}.
In these two cases, SOC destroys the symmetry of band structure thereby enhancing SHG.
Although nearly 100$\%$ spin polarization at the Fermi level is observed in materials such as half-metal Cr$_{2}$O$_{3}$ \cite{CrO},
they possess \textit{I}, where only the odd-order harmonics are observed.
Until now, little is known about 100$\%$ spin-polarized SHG in a half-metallic ferrimagnet.

In this work, we predict a strong SHG signal from the half-metallic Heusler Mn$_{2}$RuGa.
We show that a single spin channel mainly contributes to SHG in Mn$_{2}$RuGa.
The band gap in the spin-down channel is open and limits SHG, resulting in
SHG mainly from the spin-up channel.
Surprisingly, $\chi_{xyz}^{(2)}$ reaches as large as 3614 pm/V for the spin-up channel,
at least an order of magnitude larger than that of GaAs.
It is found that the intraband current dominates this large $\chi_{xyz}^{(2)}$, which originates from three highly dispersive bands near the Fermi level.
To confirm our conclusion, we remove the Ru atoms to obtain Mn$_{2}$Ga, where both the spin-up and spin-down channels are metallic.
It is found that a highly dispersive band near the Fermi level appears in the spin-down channel, which does enhance the SHG spectrum $\chi_{xyz}^{(2)}$.
With SOC, the spin-up and spin-down channels are coupled.
As a result, the restricted transitions in the spin-polarized case are now SOC-allowed between the flat valence and conduction bands.
The underlying physics stems from the reduced magnetic point group induced by SOC, where the magnetization field is applied along the $z$-axis.
This directly leads to the appearance of additional SHG components such as $\chi_{xxz}^{(2)}$, where the allowed interband transitions play a role.
Our study demonstrates that nearly 100$\%$
spin-polarized SHG can detect the half-metallicity in Heusler alloy Mn$_{2}$RuGa.

The rest of the paper is arranged as follows.
In Sec. II, we show our theoretical methods.
Then, the results and discussions are given in Sec. III.
Finally, we conclude our work in Sec. IV.

\section{Computational methods}

\subsection{First-principle electronic structure calculations }
The electronic structures of Mn$_{2}$RuGa are calculated within the first-principle density functional theory
using the projector-augmented wave (PAW) \cite{PAW1,PAW2} method,
as implemented in the Vienna Ab intio Simulation Package (VASP) \cite{vasp1,vasp2,vasp3,vasp4}.
The generalized gradient approximation (GGA) \cite{GGA} is employed within the Perdew-Burke-Ernzerhof (PBE)
scheme as the exchange-correlation functional.
We self-consistently solve the Kohn-Sham equation
\be
\begin{aligned}
&\left[-\frac{\hbar^{2}}{2m_{e}}\nabla^{2}+V_{ne}(\vec{r})+\frac{e^{2}}{4\pi\epsilon_{0}}\int\frac{n(\vec{r})}{|\vec{r}-\vec{r}'|}d^{3}\vec{r}'+V_{xc}(\vec{r})\right]\\
&\times\psi_{n\vec{k}}(\vec{r})=\varepsilon_{n\vec{k}}\psi_{n\vec{k}}(\vec{r}).
\end{aligned}
\ee
The first term is the kinetic energy and next three terms are the potential energy, the Coulomb, and the exchange interactions, respectively.
$m_{e}$ is the electron mass and $n(\vec{r})$ is the electron density.
$\psi_{n\vec{k}}(\vec{r})$ denotes the Bloch wave function of band
$n$ at crystal momentum $\vec{k}$, and $\varepsilon_{n\vec{k}}$ is the band energy.
The cutoff energy is set to 500 eV.
The structural optimizations and self-consistent are carried out
by $\Gamma$-centered $k$-point mesh of 15$\times$15$\times$15.
The density of states (DOS) is calculated
using a denser $k$-point mesh of 21$\times$21$\times$21.
In order to obtain accurate results,
we set the energy convergence less than 10$^{-6}$ eV.

\subsection{First-principle nonlinear optical response calculations}
We use the length gauge to compute SHG \cite{X1,X2,X3}.
The nonlinear polarization is given by
$P^{a}(2\omega)=\chi^{abc}(2\omega;\omega,\omega)E^{b}(\omega)E^{c}(\omega)$,
where $\chi^{abc}$ denotes the SHG susceptibility,
and $E^{b}(\omega)$ is the $b$ component of the optical electric field at frequency $\omega$.
In general, $\chi^{abc}$ contains three
major contributions: the interband transitions $\chi^{abc}_{inter}$,
the intraband transitions $\chi^{abc}_{intra}$,
and the modulation of interband terms by intraband terms
$\chi^{abc}_{mod}$, and can be expressed as
\be
\begin{aligned}
\chi^{abc}(2\omega;\omega,\omega)=&\chi^{abc}_{inter}(\omega)+
\chi^{abc}_{intra}(\omega)+\chi^{abc}_{mod}(\omega), \\
=&\chi^{abc}_{2ph,inter}(\omega)+\chi^{abc}_{1ph,inter}(\omega)  \\
&+\chi^{abc}_{2ph,intra}(\omega)+\chi^{abc}_{1ph,intra}(\omega) \\
&+\chi^{abc}_{mod}(\omega),
\end{aligned}
\ee
where the subscripts $2ph$ and $1ph$ represent two- and one-photon
transitions, respectively. The interband and intraband transitions are
\be
\begin{aligned}
\chi^{abc}_{inter}(\omega) &=\chi^{abc}_{2ph,inter}(\omega)+\chi^{abc}_{1ph,inter}(\omega),\\
\chi^{abc}_{intra}(\omega) &=\chi^{abc}_{2ph,intra}(\omega)+\chi^{abc}_{1ph,intra}(\omega).
\end{aligned}
\ee
The detailed expressions \cite{expression} of these four terms are given by
\be
\begin{aligned}
\chi^{abc}_{2ph,inter}(\omega)=&\frac{e^{3}}{\hbar^{2}}\int\frac{d\textbf{k}}{4\pi^{3}}\sum_{nml}
\frac{r_{nm}^{a}(\textbf{k})\left\{r_{ml}^{b}(\textbf{k})r_{ln}^{c}(\textbf{k})\right\}}
{\omega_{ln}(\textbf{k})-\omega_{ml}(\textbf{k})}\\
&\times\frac{2f_{nm}}{\omega_{mn}(\textbf{k})-2\omega-2i\eta},
\end{aligned}
\ee
\be
\begin{aligned}
\chi^{abc}_{1ph,inter}(\omega)=&\frac{e^{3}}{\hbar^{2}}\int\frac{d\textbf{k}}{4\pi^{3}}\sum_{nml}
\frac{r_{nm}^{a}(\textbf{k})\left\{r_{ml}^{b}(\textbf{k})r_{ln}^{c}(\textbf{k})\right\}}{\omega_{ln}(\textbf{k})-\omega_{ml}(\textbf{k})}\\
&\times\left\{\frac{f_{ml}}{\omega_{ml}(\textbf{k})-\omega-i\eta}+\frac{f_{ln}}{\omega_{ln}(\textbf{k})-\omega-i\eta}\right\},
\end{aligned}
\ee
\be
\begin{aligned}
\chi^{abc}_{2ph,intra}(\omega)=&\frac{e^{3}}{\hbar^{2}}\int\frac{d\textbf{k}}{4\pi^{3}}\sum_{nm}
\frac{r_{nm}^{a}(\textbf{k})\left\{\Delta_{mn}^{b}(\textbf{k})r_{mn}^{c}(\textbf{k})\right\}}{\omega_{mn}^{2}(\textbf{k})}\\
&\times\frac{-8if_{nm}}{\omega_{mn}(\textbf{k})-2\omega-2i\eta} \\
&-\frac{e^{3}}{\hbar^{2}}\int\frac{d\textbf{k}}{4\pi^{3}}\sum_{nml}\frac{r_{nm}^{a}(\textbf{k})\left\{r_{ml}^{b}(\textbf{k})r_{ln}^{c}(\textbf{k})\right\}}{\omega_{mn}^{2}(\textbf{k})}\\
&\times\frac{2f_{nm}\left(\omega_{ln}(\textbf{k})-\omega_{ml}(\textbf{k})\right)}{\omega_{mn}(\textbf{k})-2\omega-2i\eta},
\end{aligned}
\ee
\be
\begin{aligned}
\chi^{abc}_{1ph,intra}(\omega)=&\frac{e^{3}}{\hbar^{2}}\int\frac{d\textbf{k}}{4\pi^{3}}\sum_{nml}
r_{nm}^{a}(\textbf{k})\left\{r_{ml}^{b}(\textbf{k})r_{ln}^{c}(\textbf{k})\right\}\\
&\times\omega_{mn}(\textbf{k})\bigg\{\frac{f_{nl}}{\omega_{ln}^{2}(\textbf{k})\left(\omega_{ln}(\textbf{k})-\omega-i\eta\right)} \\
&-\frac{f_{lm}}{\omega_{ml}^{2}(\textbf{k})\left(\omega_{ml}(\textbf{k})-\omega-i\eta\right)}\bigg\}.
\end{aligned}
\ee
We know that $\chi^{abc}_{mod}$ in Eq. (2) contributes little to SHG in comparison with $\chi^{abc}_{inter}$ and $\chi^{abc}_{intra}$,
and it has the form as
\be
\begin{aligned}
\frac{i}{2\omega}\chi^{abc}_{mod}&(2\omega;\omega,\omega)=\\
\frac{ie^{3}}{2\hbar^{2}}&\int\frac{d\textbf{k}}{4\pi^{3}}\sum_{nml}
\frac{\omega_{nl}(\textbf{k})r_{lm}^{a}(\textbf{k})\left\{r_{mn}^{b}(\textbf{k})r_{nl}^{c}(\textbf{k})\right\}f_{nm}}{\omega_{mn}^{2}(\textbf{k})\left(\omega_{mn}(\textbf{k})-\omega-i\eta\right)}\\
-\frac{ie^{3}}{2\hbar^{2}}&\int\frac{d\textbf{k}}{4\pi^{3}}\sum_{nml}\frac{\omega_{lm}(\textbf{k})r_{nl}^{a}(\textbf{k})\left\{r_{lm}^{b}(\textbf{k})r_{mn}^{c}(\textbf{k})\right\}f_{nm}}{\omega_{mn}^{2}(\textbf{k})\left(\omega_{mn}(\textbf{k})-\omega-i\eta\right)}\\
+\frac{ie^{3}}{2\hbar^{2}}&\int\frac{d\textbf{k}}{4\pi^{3}}\sum_{nm}\frac{f_{nm}\Delta_{nm}^{a}(\textbf{k})\left\{r_{mn}^{b}(\textbf{k})r_{nm}^{c}(\textbf{k})\right\}}
{\omega_{mn}^{2}(\textbf{k})\left(\omega_{mn}(\textbf{k})-\omega-i\eta\right)}.
\end{aligned}
\ee
In Eqs. (4)-(8), $\omega_{mn}(\textbf{k})=\omega_{m}(\textbf{k})-\omega_{n}(\textbf{k})$, and the energy of band $n$ is $\hbar\omega_{n}$.
$\Delta_{mn}^{a}(\textbf{k})=\upsilon_{mm}^{a}(\textbf{k})-\upsilon_{nn}^{a}(\textbf{k})$, where $\upsilon_{nm}^{a}(\textbf{k})$
is the $a$ component of the velocity matrix elements, and $f_{mn}= f(\hbar\omega_{m})-f(\hbar\omega_{n})$, where $f(\hbar\omega_{n})$ is the Fermi Dirac function.
The matrix elements of position operator $r_{mn}(\textbf{k})$ is given by $r_{mn}^{a}(\textbf{k})=-i\upsilon_{nm}^{a}(\textbf{k})/\omega_{mn}(\textbf{k})$,
and $\left\{r_{ml}^{b}(\textbf{k})r_{ln}^{c}(\textbf{k})\right\}=\frac{1}{2}\left[r_{ml}^{b}(\textbf{k})r_{ln}^{c}(\textbf{k})+r_{ml}^{c}(\textbf{k})r_{ln}^{b}(\textbf{k})\right]$.
The matrix elements of position operator $r_{mn}(\textbf{k})$ are directly computed by the first-principle calculations,
which accounts for the effect of non-local potentials \cite{potential}.
The damping parameter $\eta$ is set to 0.003 Hartree.
In the realistic calculations, the number of $k$ points and energy bands affect the accuracy of $\chi^{abc}$.
A large number of $k$ points are required to obtain an accurate NLO response,
so a very dense $k$-point mesh of $32\times32\times32$ is used.
The number of energy bands is set to 32 to converge the spin-polarized SHG spectra,
and 64 for the SHG spectra with SOC.

\section{Results and discussions}

\subsection{Crystal, electronic structures and spin-polarized SHG of Mn$_{2}$RuGa }

\begin{figure}[bh]
\centering
\includegraphics[width=.45\textwidth]{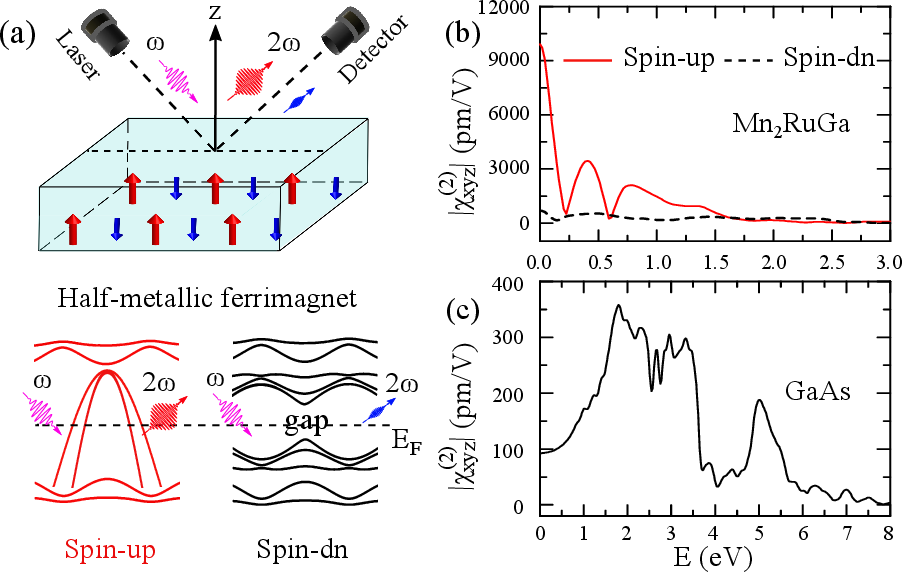}
\caption{(a) A sketch of the nearly 100$\%$ spin-polarized SHG in half-metallic ferrimagnets,
where the contribution from the spin-up channel is significant, while that from the spin-down channel is largely suppressed.
(b) The absolute value of the SHG susceptibility $\chi_{xyz}^{(2)}$ for Mn$_{2}$RuGa without SOC.
The red solid and black dashed lines denote the spin-up and spin-down channels, respectively.
The damping parameter $\eta=0.003$ Hartree is taken.
(c) $\chi_{xyz}^{(2)}$ of GaAs is also given for comparison.
}
\label{fig1}
\end{figure}
SHG appears in nonmagnetic materials with broken \textit{I}, but it does not distinguish spin.
For ferrimagnetic materials, both spin-up and spin-down channels contribute to SHG.
However, SHG from two spin channels is different, which is closely related to the spin-polarized band structures near the Fermi level.
For a half-metallic ferrimagnet with nearly 100$\%$ spin polarization at the Fermi level,
SHG mainly comes from one spin channel showing metallicity, as schematically displayed in Fig. 1(a).
Mn$_{2}$RuGa is such a half-metallic ferrimagnet with many different structures \cite{MnRuGa1,MnRuGa2,MnRuGa3}.
We choose the most stable $XA$ Heusler structure, which belongs to the
space group F-43m \cite{MnRuGa4}, as shown in Appendix A.
This structure is more consistent with the experimental results \cite{experiment}.
The experimental lattice parameters are $a=b=c=5.97$ {\AA} \cite{lattice}.
In Mn$_{2}$RuGa, there are two Mn atoms, Mn$_{1}$ and Mn$_{2}$,
with different Wyckoff positions 4a(0, 0, 0) and 4c(1/4, 1/4, 1/4).
The positions of Ru and Ga are 4d(3/4, 3/4, 3/4) and 4b(1/2, 1/2, 1/2), respectively.
The magnetic moments of two Mn atoms are $M_{4a}=3.13$ $\mu_{B}$,
and $M_{4c}=-2.29$ $\mu_{B}$, consistent with the prior study \cite{MnRuGa4}, and they are antiferromagnetically coupled.
Here, the magnetic moment direction is along the $z$ direction.
It is found that the total magnetic moment is $M_{tot}=1.03$ $\mu_{B}$,
which satisfies the Slater-Pauling rule as reported in the literature \cite{SP}.
This rule provides a simple relationship between the total
magnetic moment $M_{tot}$ and the valence electron $Z$.
For Mn$_{2}$RuGa, they satisfy $M_{tot}=Z-24$,
and the number of valence electron is 25,
so $M_{tot}$ is about 1 $\mu_{B}$ \cite{SP-MnRuGa}.

\begin{table}[bt]
\centering
\caption{The SHG susceptibilities for different materials.
$|\chi^{(2)}|$ and $|\chi^{(2)}|_{SP}$ represent the absolute values of SHG susceptibilities without and with spin polarization, respectively, in units of pm/V.
The unit of the photon energy is eV. cMQWs is the abbreviation of the coupled metallic quantum wells.
}
\begin{tabular}{cccccccccccccccccccc}\hline\hline
Material                  &&& $|\chi^{(2)}|$  &&& $|\chi^{(2)}|_{SP}$ &&&  Photon energy   &&& Reference                 \\ \hline
Mn$_{2}$RuGa              &&&                 &&&3614                 &&&0.38           &&& This work                   \\
GaAs                      &&& 358             &&&                     &&&1.80            &&& This work                    \\
GaAs                      &&& 350             &&&                     &&&1.53            &&& Ref.\cite{table1}             \\
TaAs                      &&& 3600            &&&                     &&&1.55            &&& Ref.\cite{table2}              \\
Co$_{3}$Sn$_{2}$S$_{2}$   &&& 10$^{5}$        &&&                     &&&0.05          &&& Ref.\cite{table3}               \\
cMQWs                     &&& 1500            &&&                     &&&1.35            &&& Ref.\cite{table4}                \\
BiFeO$_{3}$               &&& 15-19           &&&                     &&&0.80          &&& Ref.\cite{table5}                 \\
CaCoSO                    &&&                 &&&6.9                  &&&1.17          &&& Ref.\cite{table6}                  \\
\hline\hline
\end{tabular}\label{tab1}
\end{table}
In order to understand the nonlinear optical properties of Mn$_{2}$RuGa,
we have calculated the second-order nonlinear optical susceptibilities.
Nonmagnetic Mn$_{2}$RuGa belongs to the point group T$_{d}$.
There are six equivalent nonvanishing SHG susceptibilities
$\chi_{xyz}^{(2)}=\chi_{xzy}^{(2)}=\chi_{yxz}^{(2)}=\chi_{yzx}^{(2)}=\chi_{zxy}^{(2)}=\chi_{zyx}^{(2)}$.
When the antiferromagnetic coupling appears along the $z$ direction,
the symmetry is reduced from 24 to 8, belonging to the magnetic point group -42m.
This changes SHG. The system contains three independent nonvanishing elements, namely,
$\chi_{xyz}^{(2)}=\chi_{yxz}^{(2)}$, $\chi_{xzy}^{(2)}=\chi_{yzx}^{(2)}$, and $\chi_{zxy}^{(2)}=\chi_{zyx}^{(2)}$.
The SHG susceptibility satisfies the intrinsic permutation symmetry, that is $\chi_{abc}^{(2)}=\chi_{acb}^{(2)}$.
Thus, there are only two independent nonvanishing elements, $\chi_{xyz}^{(2)}=\chi_{xzy}^{(2)}=\chi_{yxz}^{(2)}=\chi_{yzx}^{(2)}$,
and $\chi_{zxy}^{(2)}=\chi_{zyx}^{(2)}$.

For the SHG susceptibility $\chi_{xyz}^{(2)}$, the absolute value of the spin-up channel
is given by the red solid line in Fig. 1(b).
It is shown that $|\chi_{xyz}^{(2)}|$ is at the maximum value
of 11670.76 pm/V when the photon energy approaches 0 eV.
Due to the metallic nature for the spin-up channel in Mn$_{2}$RuGa, the intensity of $|\chi_{xyz}^{(2)}|$ at 0 eV is not accurate,
which also depends on the damping parameter [see Appendix A for more details].
When the photon energy is between 0 and 0.22 eV, $|\chi_{xyz}^{(2)}|$ decreases monotonically.
At 0.22 eV, the value of $|\chi_{xyz}^{(2)}|$ is close to zero. As the photon energy increases,
a dramatic peak appears at 0.38 eV, as shown in Table \uppercase\expandafter{\romannumeral1}.
The intensity of this peak is as large as 3614.37 pm/V.
We also note that this peak is insensitive to the damping parameter, as discussed in Appendix A.
As the energy further increases, the spectrum oscillates and gradually decreases,
and the final intensity is close to zero. By contrast, the spin-down $|\chi_{xyz}^{(2)}|$
[see the black dashed line in Fig. 1(b)] is much smaller, with a maximum of about 652.46 pm/V in the entire energy range.
This is consistent with the experimental finding \cite{SHG-experiment},
where only the majority spin channel is optically excited highly.
When the energy is less than 0.2 eV, the intensity difference of
$|\chi_{xyz}^{(2)}|$ between the spin-down and spin-up channels is largest.
In the energy range of 0.2 to 1.6 eV, the spin-down $|\chi_{xyz}^{(2)}|$ change is relatively stable,
but the spin-up decreases, so the difference between the two spins decreases.
When the energy is larger than 1.6 eV, the spin-down $|\chi_{xyz}^{(2)}|$ is very close to the spin-up,
and the intensity is almost zero.
Therefore, we can conclude that SHG in Mn$_{2}$RuGa is mainly contributed by the spin-up channel \cite{SHG-experiment},
while the contribution of spin-down channel is negligible.
As a result, we obtain nearly 100$\%$ spin-polarized SHG in Mn$_{2}$RuGa.
We also notice that GaAs and Mn$_{2}$RuGa share the same point group.
However, as shown in Fig. 1(c), the maximum intensity of $|\chi_{xyz}^{(2)}|$ in GaAs is only about 358.19 pm/V at 1.8 eV,
which is about 10 times smaller than the spin-polarized SHG of Mn$_{2}$RuGa.
The SHG susceptibilities for other materials are also presented in Table \uppercase\expandafter{\romannumeral1}.
We believe that such a large spin-polarized SHG in Mn$_{2}$RuGa has potential applications in spin-filter devices \cite{s-f}.

\begin{figure}
\centering
\includegraphics[width=.45\textwidth]{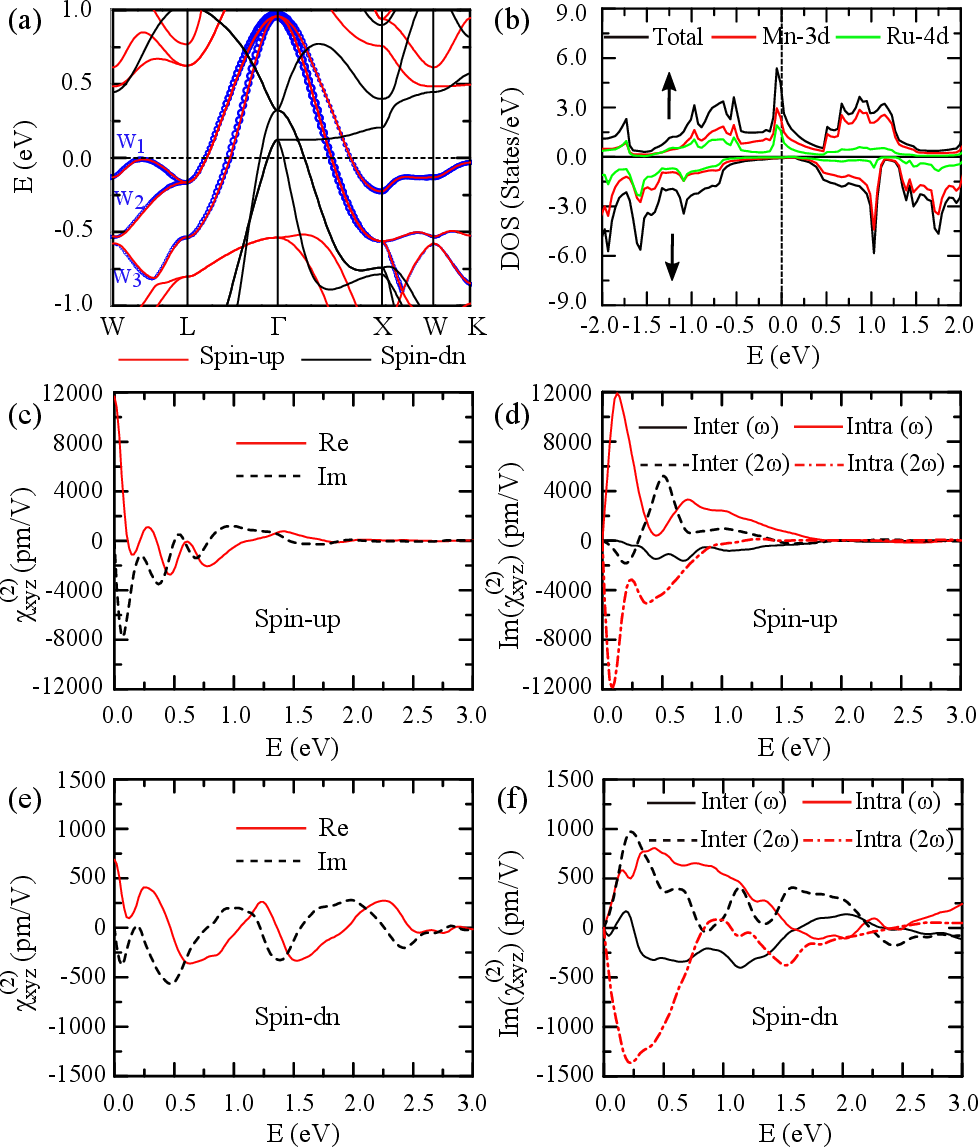}
\caption{(a) Band structure without SOC for Mn$_{2}$RuGa,
where red line represents the spin-up channel, and black line represents the spin-down channel.
The blue circles represent the $d_{xy}$, $d_{yz}$ and $d_{xz}$ orbitals of Mn atoms for the bands W$_{1}$, W$_{2}$ and W$_{3}$.
The dashed line denotes the Fermi level.
(b) Total DOS of Mn$_{2}$RuGa, where the partial DOS of Mn-$3d$ and Ru-$4d$ states are also
given and represented by red and green lines, respectively.
The upward arrow indicates the spin-up channel, and the downward arrow indicates the spin-down channel.
The vertical dashed line denotes the Fermi level.
(c) Real and imaginary parts of the SHG susceptibility $\chi_{xyz}^{(2)}$ from the spin-up channel in Mn$_{2}$RuGa without SOC.
(d) Calculated Im($\chi_{xyz}^{(2)}$) from inter($\omega$)/(2$\omega$) (black solid, black dashed curve)
and intra ($\omega$)/(2$\omega$) (red solid, red dashed-dotted curve) parts.
(e) and (f) are similar to (c) and (d), but from the spin-down channel.
}
\label{fig2}
\end{figure}

\subsection{Intraband and interband contributions}
To reveal insights into this nearly 100$\%$ spin-polarized SHG, we resort to the band structure of Mn$_{2}$RuGa, as shown in Fig. 2(a).
The red line indicates the spin-up channel, and the black line indicates the spin-down channel.
The band of spin-up channel crosses the Fermi level at multiple $k$ points, indicating a metallic state.
This result can be verified from the total density of states (DOS), as shown in Fig. 2(b).
Bands in the energy range from $-0.08$ to 0.12 eV are mainly occupied by $d$ orbitals,
in which Mn-3$d$ orbitals are dominant, Ru-4$d$ orbitals contribute less,
and the contribution from Ga-3$d$ orbitals is negligible.
However, the spin-down band only touches the Fermi level near the $\Gamma$ point.
The total DOS near the Fermi level is close to 0.
Therefore, Mn$_{2}$RuGa is a half metal, which is consistent with previous report \cite{MnRuGa3,MnRuGa4}.
The band gap disappears in the spin-up channel, but appears in the spin-down channel.
Therefore, the appearance of band gap hinders the transition of electrons from the valence band to the conduction band.

Next, we further analyze the difference between the spin-up and the spin-down $\chi_{xyz}^{(2)}$ of Mn$_{2}$RuGa
by examining the real and imaginary parts separately.
For the spin-up channel, the real and imaginary parts of $\chi_{xyz}^{(2)}$
are given in red solid and black dashed lines, respectively, as shown in Fig. 2(c).
The real part of $\chi_{xyz}^{(2)}$ (Re($\chi_{xyz}^{(2)}$)) decreases monotonously
as the photon energy increases from 0 to 0.16 eV.
When the photon energy reaches 0.16 eV, its value reaches a minimum of -1099.71 pm/V.
When the energy is larger than 0.16 eV, the spectrum oscillates and finally approaches zero.
For the imaginary part of $\chi_{xyz}^{(2)}$, that is Im($\chi_{xyz}^{(2)}$), a Lorentzian-like resonance appears
in the energy range between 0 and 0.4 eV.
Compared with the real part, its first negative peak shifts to the lower energy by about 0.06 eV,
and the intensity is decreased to 7768.47 pm/V.
When the photon energy further increases, the spectrum oscillates.

In general, the imaginary part of $\chi_{xyz}^{(2)}$ reflects the optical absorption in NLO experiments.
From the band structure, the optical absorption involves in the intra- and interband currents.
Thus, we decompose Im($\chi_{xyz}^{(2)}$) into the inter- and intraband parts for the spin-up channel, as shown in Fig. 2(d).
It clearly shows that the intraband contribution dominates the spectrum in the lower photon energy window from 0 to 0.4 eV.
By contrast, the interband contribution from both single- and two-photon resonances is much smaller.
In the energy range of around 0.4$-$2.0 eV, these four spectra are comparable and have opposite signs for the single- and two-photon resonances,
leading to the oscillation of Im($\chi_{xyz}^{(2)}$) in this energy range.
Thus, the negative characteristic peak of Im($\chi_{xyz}^{(2)}$) in the lower energy is determined by the intraband current.
In other words, the interband current contributes little to the negative characteristic peak in the lower energy.
We know that the negative characteristic peak locates in the energy range of 0$-$0.4 eV.
The two-photon resonance would correspond to the energy range of around 0$-$0.8 eV.
In this perspective, the bands related to the intraband current would locate in the energy range from $-0.4$ to 0.4 eV near the Fermi level.
As shown in Fig. 2(a), we can see that only three bands W$_{1}$, W$_{2}$ and W$_{3}$ appear in this energy range.
The bands W$_{1}$ and W$_{2}$ are degenerate along the $L$-$\Gamma$ direction,
while the bands W$_{2}$ and W$_{3}$ are degenerate along the $\Gamma$-$X$ direction.
More importantly, these three bands disperse quadratically along the $L$-$\Gamma$-$X$ direction.
This means that they highly disperse along this high-symmetry line,
contributing a large normal velocity to the intraband current \cite{Niu}.
This is the reason why the intraband current dominates the negative characteristic peak in the lower energy range.

We also notice that there exist two regions for the interband transitions among these three bands.
One is located near the $L$ point along the $L$-$\Gamma$ direction,
where the double degenerate bands of W$_{1}$ and W$_{2}$ form the conduction bands
while the band W$_{3}$ is the valence band.
The other interband transition appears near the $X$ point along the $\Gamma$-$X$ direction,
where the band W$_{1}$ is the conduction band while the two-fold degenerate bands of W$_{2}$ and W$_{3}$ are the valence bands.
According to the selection rules, the interband transitions from these two regions are largely limited.
This is because these three bands are mainly formed by the $d_{xy}$, $d_{yz}$ and $d_{xz}$
orbitals of Mn atoms [see Fig. 2(a) for more details].
The interband transitions between the same $d$ orbitals are not allowed.
As a result, the interband current contributes little to the negative characteristic peak in the lower energy range.
This is generic for metallic ferro- or ferrimagnets as the spin-splitting states near the Fermi level are dominated by the $d$ orbitals.

In the case of spin-down channel, the real and imaginary parts of $\chi_{xyz}^{(2)}$ are comparable, as shown in Fig. 2(e).
Both oscillate around zero as the photon energy increases.
However, they are largely suppressed in comparison with those of spin-up channel.
Similarly, we also decompose Im($\chi_{xyz}^{(2)}$) into the inter- and intraband contributions, as shown in Fig. 2(f).
It is found that both the inter- and intraband currents are comparable and oscillate around zero,
which are much smaller than those of the spin-up channel.
This is because only few electrons are allowed to transit from the valence band to the conduction band near the $\Gamma$ point.
However, the large band gap of the spin-down channel limits both the inter- and intraband currents.
This explains why SHG of the spin-down channel is much smaller.

\subsection{SHG in metallic Mn$_{2}$Ga}
To further confirm the contribution of SHG from the spin-up quadratic bands in Mn$_{2}$RuGa,
we artificially remove the Ru atoms in Mn$_{2}$RuGa and obtain the crystal structure of Mn$_{2}$Ga \cite{MnRuGa4}.
In Mn$_{2}$Ga, two Mn atoms are antiferromagnetic coupled
and their magnetic moments are close to 3 $\mu_{B}$, which can compensate each other.
The magnetic moments of Ga atoms are very small,
so the total magnetic moment of the unit cell is nearly zero.
This coincides with the previous report \cite{SP-MnRuGa}.
However, the antiferromagnetic coupling in Mn$_{2}$Ga has a huge effect on the spin-polarized band structures, as shown in Fig. 3(a).
It is found that both the spin-up and spin-down bands cross the Fermi level,
indicating that Mn$_{2}$Ga is a metal. This is also confirmed from the PDOS, as shown in Fig. 3(b).
We can see that obvious DOS exits both for the spin-up and spin-down channels near the Fermi level,
where the Mn-$3d$ orbitals dominates.
\begin{figure}[bh]
\centering
\includegraphics[width=.45\textwidth]{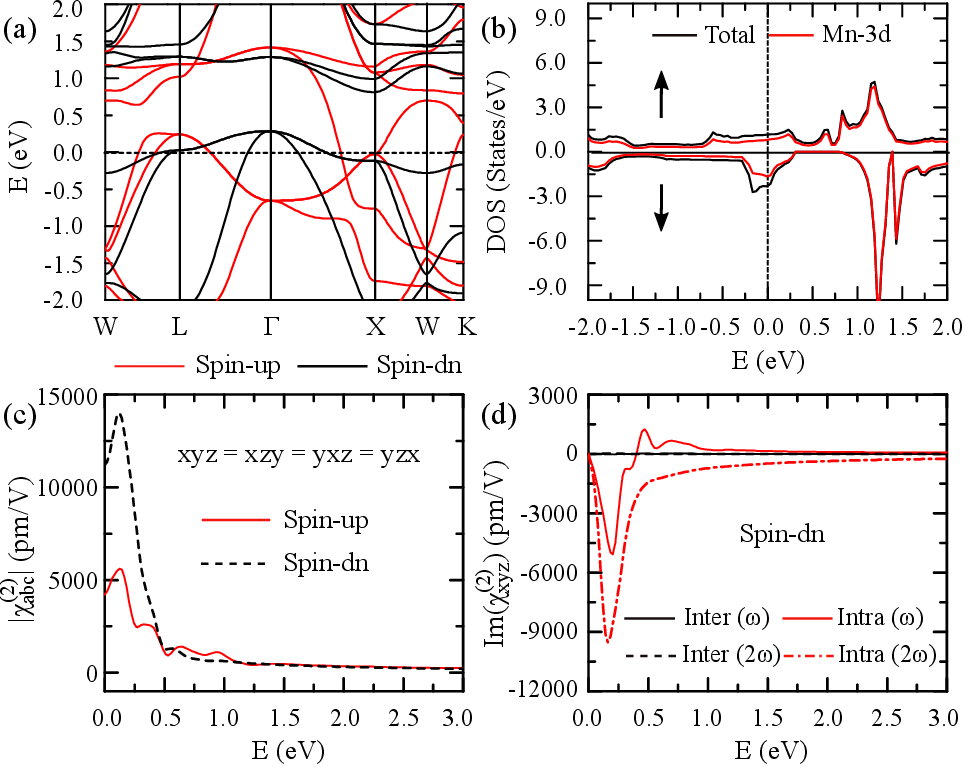}
\caption{(a) Band structures for Mn$_{2}$Ga, where the red and black lines represent the spin-up and spin-down channels, respectively.
The dashed line denotes the Fermi level.
(b) Total DOS of Mn$_{2}$Ga, where the partial DOS of Mn-$3d$ orbitals is also
given and represented by red line.
The upward arrow indicates the spin-up channel, and the downward arrow indicates the spin-down channel.
The vertical dashed line denotes the Fermi level.
(c) The absolute value of the SHG susceptibility $\chi_{abc}^{(2)}$ for Mn$_{2}$Ga, where $abc$ refers to $xyz$, $xzy$, $yxz$, and $yzx$.
The red solid and black dashed lines denote the spin-up and spin-down channels, respectively.
(d) Calculated Im($\chi_{xyz}^{(2)}$) from inter($\omega$)/(2$\omega$) (black solid, black dashed curve)
and intra ($\omega$)/(2$\omega$) (red solid, red dashed-dotted curve) parts for the spin-down channel.
}
\label{fig3}
\end{figure}

For Mn$_{2}$Ga, SHG has two independent nonvanishing elements, namely,
$\chi_{xyz}^{(2)}=\chi_{xzy}^{(2)}=\chi_{yxz}^{(2)}=\chi_{yzx}^{(2)}$,
and $\chi_{zxy}^{(2)}=\chi_{zyx}^{(2)}$.
The absolute value of $\chi_{xyz}^{(2)}$ is displayed in Fig. 3(c).
For the spin-up channel, the first peak appears at the photon energy 0.14 eV and its intensity is 5602.48 pm/V.
Then, the intensity sharply decreases and finally approaches zero as the photon energy further increases.
In the case of the spin-down channel, $\chi_{xyz}^{(2)}$ is similar to that of the spin-up channel,
but the intensity is much larger.
However, this is contrast to that of the spin-down channel in Mn$_{2}$RuGa, where $\chi_{xyz}^{(2)}$ is much smaller.
This is because no band gap appears in the spin-down channel of Mn$_{2}$Ga and three metallic bands cross the Fermi level.
More importantly, one of them disperses quadratically with $k$ near the $\Gamma$ point, which largely contributes to the intraband part of $\chi_{xyz}^{(2)}$.
This implies that the presence of band gap in the spin-down channel of Mn$_{2}$RuGa limits $\chi_{xyz}^{(2)}$,
while the absence of band gap or the quadratic band near the Fermi level enhances $\chi_{xyz}^{(2)}$.
To this end, we decompose Im($\chi_{xyz}^{(2)}$) of the spin-down channel into the inter- and intraband contributions, as shown in Fig. 3(d).
It clearly shows that the intraband current dominates, while the interband contribution is nearly neglected.
In addition, the two-photon resonance is obviously larger than that of one-photon resonance.
This would be easily detected by SHG in experiment.

\subsection{Role of SOC in SHG in Mn$_{2}$RuGa and the group symmetry}
\begin{figure}
\centering
\includegraphics[width=.45\textwidth]{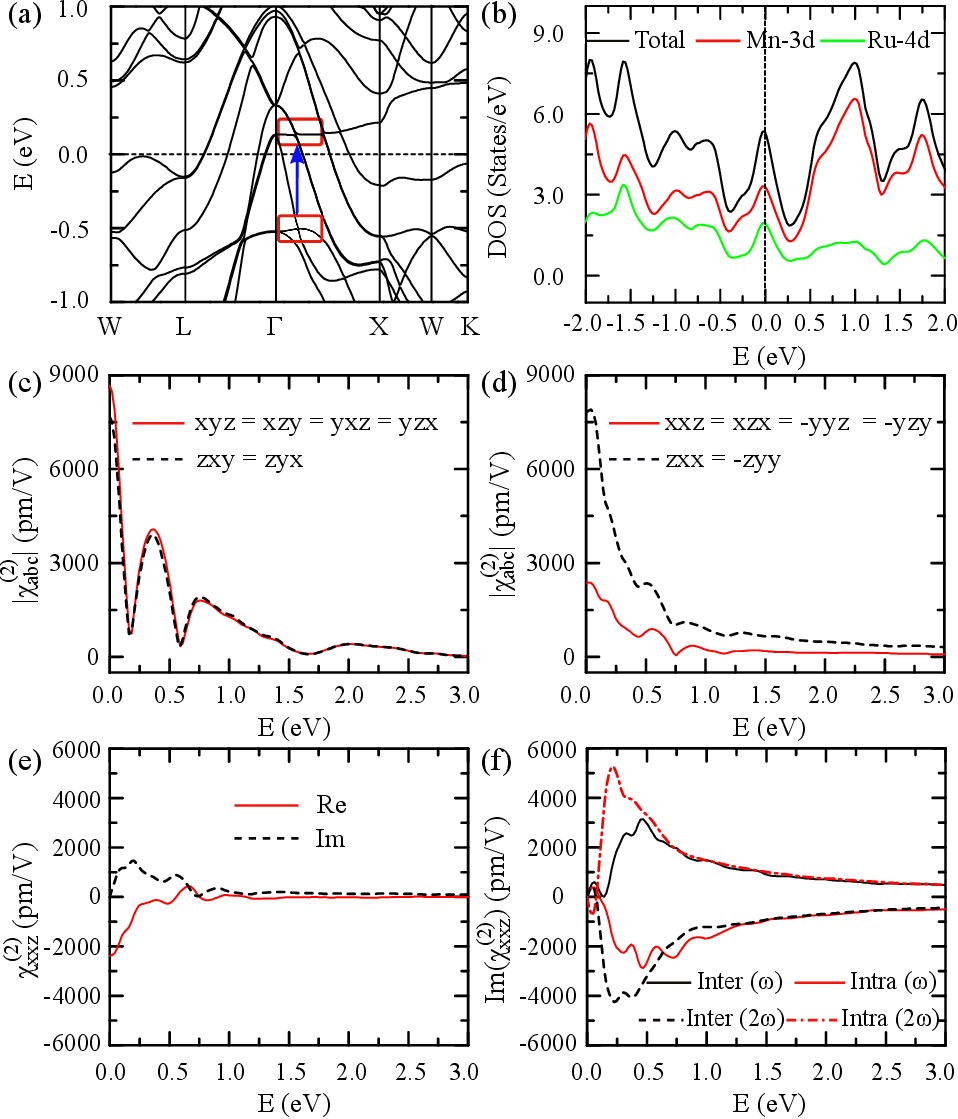}
\caption{ (a) Band structure for Mn$_{2}$RuGa under SOC.
The dashed line denotes the Fermi level.
The rectangles denote the flat valence and conduction bands and the arrow labels the transitions between those flat bands.
(b) Total DOS of Mn$_{2}$RuGa, where the partial DOS of Mn-$3d$ and Ru-$4d$ states are also
given and represented by red and green lines, respectively.
The vertical dashed line denotes the Fermi level.
(c) and (d) The absolute values of the SHG susceptibility $\chi_{abc}^{(2)}$ for Mn$_{2}$RuGa under SOC.
(e) Real and imaginary parts of $\chi_{xxz}^{(2)}$.
(f) Calculated Im($\chi_{xxz}^{(2)}$) from inter($\omega$)/(2$\omega$) (black solid, black dashed curve)
and intra ($\omega$)/(2$\omega$) (red solid, red dashed-dotted curve) parts.
}
\label{fig4}
\end{figure}
With SOC, the spin-up and the spin-down channels mix together.
The band structure of Mn$_{2}$RuGa is shown in Fig. 4(a),
which almost coincides with the spin-polarized band structures.
Figure 4(b) shows that DOS has a peak between $-0.1$ and 0.1 eV,
and mainly comes from the Mn-$3d$ orbitals, which is the same as the spin-up channel.
This is because the spin-down DOS is close to zero in this energy range.
However, in the energy range below $-0.1$ eV and above 0.1 eV,
DOS under SOC changes compared to the spin-up channel,
which is due to the contribution of the spin-down channel.
This will affect SHG.

In fact, if we only include SOC, but ignore the magnetic field direction, the symmetries of the system remain unchanged.
In real calculations, SOC is considered through a tiny magnetic field applied along the $z$-axis.
The symmetry is reduced and belongs to the magnetic point group of -4.
The remaining four symmetry operations are the identity operation $E$,
the twofold rotational symmetry $C_{2z}$ with the binary axis as
the $z$ axis, and two combination operations $IC_{4z}$ and $IC^{-1}_{4z}$.
$IC_{4z}$ denotes the rotation of $\pi$/2 around the $z$-axis,
followed by a mirror symmetry $\sigma_{xy}$.
$IC^{-1}_{4z}$ is similar to $IC_{4z}$, but with a rotation $-\pi$/2 around the $z$-axis.
As a result, there are six independent nonvanishing elements, $\chi_{xyz}^{(2)}=\chi_{yxz}^{(2)}$,
$\chi_{xzy}^{(2)}=\chi_{yzx}^{(2)}$, $\chi_{zxy}^{(2)}=\chi_{zyx}^{(2)}$,
$\chi_{xxz}^{(2)}=-\chi_{yyz}^{(2)}$, $\chi_{xzx}^{(2)}=-\chi_{yzy}^{(2)}$, and $\chi_{zxx}^{(2)}=-\chi_{zyy}^{(2)}$.
Based on the intrinsic permutation symmetry, only four components indeed appear,
$\chi_{xyz}^{(2)}=\chi_{xzy}^{(2)}=\chi_{yxz}^{(2)}=\chi_{yzx}^{(2)}$, $\chi_{zxy}^{(2)}=\chi_{zyx}^{(2)}$,
$\chi_{xxz}^{(2)}=\chi_{xzx}^{(2)}=-\chi_{yyz}^{(2)}=-\chi_{yzy}^{(2)}$, and $\chi_{zxx}^{(2)}=-\chi_{zyy}^{(2)}$.
It should be noted that $\chi_{xxz}^{(2)}$ and $\chi_{zxx}^{(2)}$ are induced from the reduced magnetic point group.

Figure 4(c) shows the absolute values of SHG susceptibilities $\chi_{xyz}^{(2)}$ and $\chi_{zxy}^{(2)}$.
It clearly shows that those two SHG spectra are nearly the same as that of the spin-up channel [see Fig. 1(d)].
This is because the contribution from the spin-down channel is negligible.
Thus, SOC has little effect on these six SHG spectra, which also appear in the spin-polarized case.
It is also found that the vanishing SHG spectra $\chi_{xxz}^{(2)}$ and $\chi_{zxx}^{(2)}$
in the spin-polarized case are recovered now, as shown in Fig. 4(d).
The intensity of $\chi_{zxx}^{(2)}$ is obviously larger than that of $\chi_{xxz}^{(2)}$.
The appearance of them directly comes from the reduced magnetic point group.
In the following, we use the four remaining symmetries to understand the nonvanishing SHG spectrum $\chi_{xxz}^{(2)}$.
The matrix representations of $E$ and $C_{2z}$ are diag$\{1,1,1\}$ and diag$\{-1,-1,1\}$.
The other two $IC_{4z}$ and $IC^{-1}_{4z}$ are
\be
IC_{4z}=
\begin{pmatrix}
    0 && 1 && 0 \\
   -1 && 0 && 0 \\
    0 && 0 && -1
\end{pmatrix}, \\
IC^{-1}_{4z}=
\begin{pmatrix}
    0 && -1 && 0 \\
   1 && 0 && 0 \\
    0 && 0 && -1
\end{pmatrix}.
\ee
The transformations of position operator under these four symmetry operations are as follows.
$E$: $(x,y,z)\rightarrow(x,y,z)$, $C_{2z}$: $(x,y,z)\rightarrow(-x,-y,z)$,
$IC_{4z}$: $(x,y,z)\rightarrow(y,-x,-z)$,
and $IC^{-1}_{4z}$: $(x,y,z)\rightarrow(-y,x,-z)$.
As a result, we can get:
\be
\begin{aligned}
E: \chi_{xxz}^{(2)}&\rightarrow\chi_{xxz}^{(2)},\\
C_{2z}: \chi_{xxz}^{(2)}&\rightarrow\chi_{(-x)(-x)z}^{(2)}=\chi_{xxz}^{(2)},\\
IC_{4z}: \chi_{xxz}^{(2)}&\rightarrow\chi_{yy(-z)}^{(2)}=-\chi_{yyz}^{(2)},\\
IC^{-1}_{4z}: \chi_{xxz}^{(2)}&\rightarrow\chi_{(-y)(-y)(-z)}^{(2)}=-\chi_{yyz}^{(2)}.
\end{aligned}
\ee
It clearly shows that these four SHG susceptibilities do not cancel out each other.
We can obtain these induced SHG spectra via the above symmetry analysis,
where $\chi_{xxz}^{(2)}$ and $\chi_{-yyz}^{(2)}$ are protected by $C_{2z}$ and $IC_{4z}$, respectively.

The reduced magnetic point group is closely related to SOC and the applied magnetic filed direction.
This means that the induced SHG spectra must have a deep relation to them.
Here, we take $\chi_{xxz}^{(2)}$ to reveal the underlying physics.
The decomposed real and imaginary parts of $\chi_{xxz}^{(2)}$ are displayed in Fig. 4(e).
We can see that both have a similar manner.
Sizable intensities are mainly located in the photon energy range of 0$-$0.75 eV.
To understand the contributions of inter- and intraband currents,
we decompose Im($\chi_{xxz}^{(2)}$) into the inter- and intraband parts, as shown in Fig. 4(f).
There also exists a large contribution from the intraband current, where the highly dispersed bands near the Fermi level play a role.
It should be noted that the interband contributions are largely enhanced, which are nearly limited in the spin-polarized case.
We find that many transitions are now allowed between the flat valence and conduction bands, as shown in rectangles of Fig. 4(a).
However, those transitions are not allowed in the spin-polarized case. This is because the flat valence bands are originally spin-up polarized,
while those flat conduction bands are spin-down polarized.
The direct transition from the spin-up band to the spin-down band is forbidden.
But, this does occur under SOC as the conservation of spin is not needed.
To check it, we calculate some matrix elements of the position operator between those flat bands at several $k$ points, as shown in Table \uppercase\expandafter{\romannumeral2}.
We can see that the $y$-component of matrix elements are small but not zero.
However, the $x$- and $z$-components are much larger.
The maximum absolute value reaches as large as 8.02 a$_{0}$ with a$_{0}$ being the Bohr radius.
These nonzero matrix elements confirm that the transitions are not allowed in the spin-polarized case, but did occur with SOC.
This tells us that the reduced magnetic point group or the remaining four symmetries protect the induced SHG spectra, which is similar to our previous study \cite{bise}.

\begin{table}[bh]
\centering
\caption{The matrix elements of position operator between those flat bands near the Fermi level under SOC in Mn$_{2}$RuGa for three $k$ points, where the atomic unit is used.
}
\begin{tabular}{ccccccccccccccccccccccc}\hline\hline
\multirow{2}{1.1cm}{$k$ point}  &&&\multicolumn{2}{c}{$x$}&&&\multicolumn{2}{c}{$y$}&&&\multicolumn{2}{c}{$z$}       \\ \cline{4-5}\cline{8-9}\cline{12-13}
                  &&&   Re&Im       &&&    Re&Im      &&&    Re&Im        \\ \hline
$(0,0.139,0)$     &&&-0.83&0.32     &&&  0.05&0.12    &&&  -0.68&-1.74       \\
$(0,0.209,0)$     &&& -1.07&-2.34   &&&  0.01&-0.01   &&&  -2.49&1.13        \\
$(0,0.278,0)$     &&& 1.93&2.45     &&& -0.09&0.07    &&&  6.29&-4.96         \\
\hline\hline
\end{tabular}\label{tab1}
\end{table}

\section{Conclusions}

In conclusion, we have demonstrated that large nearly 100$\%$ spin-polarized SHG carries rich
information about the electronic structures of the half-metal Mn$_{2}$RuGa.
The band gap in the spin-down channel limits SHG.
In contrast, the spin-up channel is metallic and gives rise to $\chi_{xyz}^{(2)}$ as large as 3614 pm/V,
which is about 10 times larger than that of typical nonlinear materials such as GaAs.
For the spin-up channel, the intraband current mainly contributes to $\chi_{xyz}^{(2)}$, which stems from three highly dispersed bands near the Fermi level.
In addition, SOC under the $z$-axis magnetization field induces additional SHG susceptibilities such as $\chi_{xxz}^{(2)}$ from the reduced magnetic point group, where the interband transitions dominate.
Our study would provide a good guide in future application of large spin-polarized SHG in spin-filter devices.

\section*{ACKNOWLEDGMENTS}
This work was supported by the National Science Foundation of China under Grant No. 11874189.
We also acknowledge the Fermi cluster at Lanzhou University for providing computational resources.
GPZ was supported by the U.S. Department of Energy under Contract No. DE-FG02-06ER46304.
The research used resources of the National Energy
Research Scientific Computing Center, which is supported by
the Office of Science of the U.S. Department of Energy under
Contract No. DE-AC02-05CH11231.

$^{*}$sims@lzu.edu.cn

$^{\dagger}$guo-ping.zhang@outlook.com

\subsection*{APPENDIX A: The crystal structure of Mn$_{2}$RuGa and the effect of damping parameters on SHG}
Heusler alloys have three different structures belonging to different space groups \cite{structure1}.
The normal full-Heusler $X_{2}YZ$ alloys belong to group symmetry L2$_{1}$ (No. 225).
The Half-Heusler $XYZ$ compounds have group symmetry $C1_{b}$ (No. 216), and
the inverse-Heusler $X_{2}YZ$ alloys with group symmetry $XA$ (No. 216) \cite{MnRuGa2}.
We select a stable one, that is $XA$ structure, as our example, as shown in Fig. 5(a),
where the sublattice Mn$_{2}$RuGa are ferrimagnetic ordering.
The magnetic moments of these sublattice Mn atoms are 3.13 and -2.29 $\mu_{B}$, respectively.
As a result, the net magnetic moment of unit cell is about 1 $\mu_{B}$,
which agrees well with the Slater-Pauling rule.

\begin{figure}
\centering
\includegraphics[width=.45\textwidth]{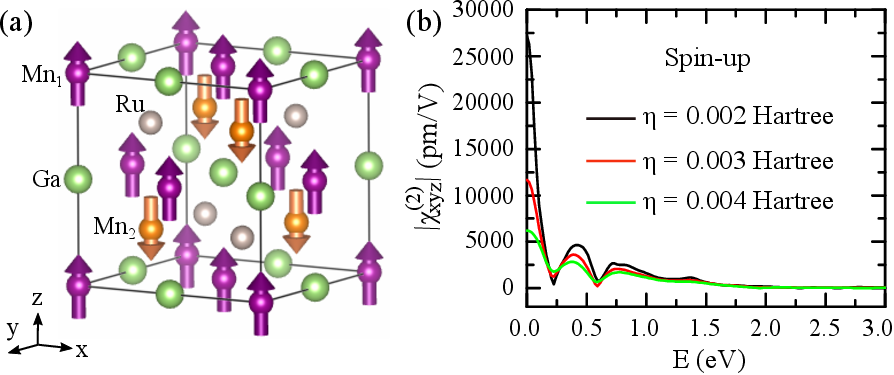}
\caption{ (a) Crystal structure of Mn$_{2}$RuGa. Purple, orange, grey, and green spheres represent Mn$_{1}$, Mn$_{2}$, Ru, and Ga atoms, respectively.
The arrows marked on the Mn atoms indicate the local magnetic moments, forming the antiferromagnetic configuration.
(b) The absolute value of the spin-up $\chi_{xyz}^{(2)}$ of Mn$_{2}$RuGa without SOC for different damping parameters $\eta$.
}
\label{fig5}
\end{figure}

Due to the metallic nature for the spin-up channel in Mn$_{2}$RuGa, the nonlinear optical response at zero frequency must be estimated,
because the energy difference $\hbar\omega_{nm}=E_{n}-E_{m}$ is zero for those metallic states near the Fermi level.
As a result, $\omega_{nm}-\omega$ or $\omega_{nm}-2\omega$ in the denominators of Eqs. (4)-(7) in the main text diverges.
To avoid this, the damping parameter $\eta$ is introduced to estimate $|\chi^{(2)}|$ at zero frequency.
However, the obtained results are still not accurate.
As shown in Fig. 5(b), we can see that the intensity of the spin-up $|\chi_{xyz}^{(2)}|$ at 0 eV largely depends on $\eta$.
When $\eta=0.004$ Hartree, the intensity of spin-up $|\chi_{xyz}^{(2)}|$ at 0 eV is about 6150.37 pm/V.
When $\eta$ is decreased to 0.003 Hartree, the value is about twice larger than that for $\eta=0.004$ Hartree.
When $\eta$ is further decreased to 0.002 Hartree, the intensity of spin-up $|\chi_{xyz}^{(2)}|$ is dramatically increased.
This means that $|\chi^{(2)}|$ usually diverges at zero frequency.
In other words, it is a challenge to accurately compute $|\chi^{(2)}|$ at zero frequency.
This is also the case of linear response in Hall effect.
According to the Drude model, the frequency-dependent conductivity $\sigma(\omega)$ is given by,
\be
\begin{aligned}
\sigma(\omega)=\frac{\sigma_{0}}{1-i\omega\tau},
\end{aligned}
\ee
where $\sigma_{0}$ is the DC Drude conductivity without the external magnetic field and $\tau$ is the relaxation time.
At zero frequency, we can see $\sigma(\omega)$ reduces to $\sigma_{0}=ne^{2}\tau/m$
with $n$, $e$, and $m$ being the electron density, the electron charge, and the electron mass, respectively.
This shows the failure of zero-frequency response already exists in the linear response, which would be our future research focus of SHG.

When the photon energy is further increased to about 0.38 eV, a stable peak of spin-up $|\chi_{xyz}^{(2)}|$ with respect to $\eta$ appears.
The intensity of this peak increases as $\eta$ decreases.
But the change is small.
This implies that the intensity of this second peak is insensitive to $\eta$.
We also note that the intensity of this peak corresponding to $\eta=0.003$ Hartree is as large as 3614.37 pm/V,
which can be compared with SHG spectra of other materials such as GaAs, TaAs, and CaCoSO [see Table I].

\subsection*{APPENDIX B: The SHG susceptibility for Mn$_{2}$RuGa without SOC}
\begin{figure}[bh]
\centering
\includegraphics[width=.25\textwidth]{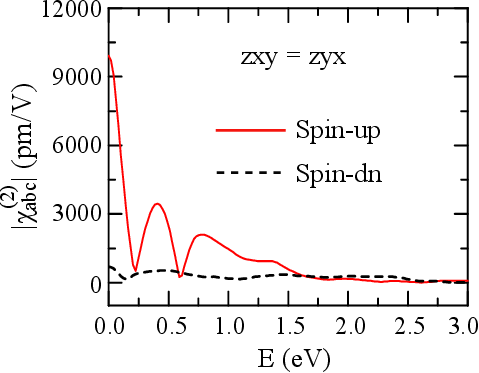}
\caption{ The absolute value of the SHG susceptibilities $\chi_{abc}^{(2)}$ for Mn$_{2}$RuGa without SOC.
The red solid and black dashed lines denote the spin-up and spin-down channels, respectively.}
\label{fig6}
\end{figure}
\begin{figure}[bh]
\centering
\includegraphics[width=.45\textwidth]{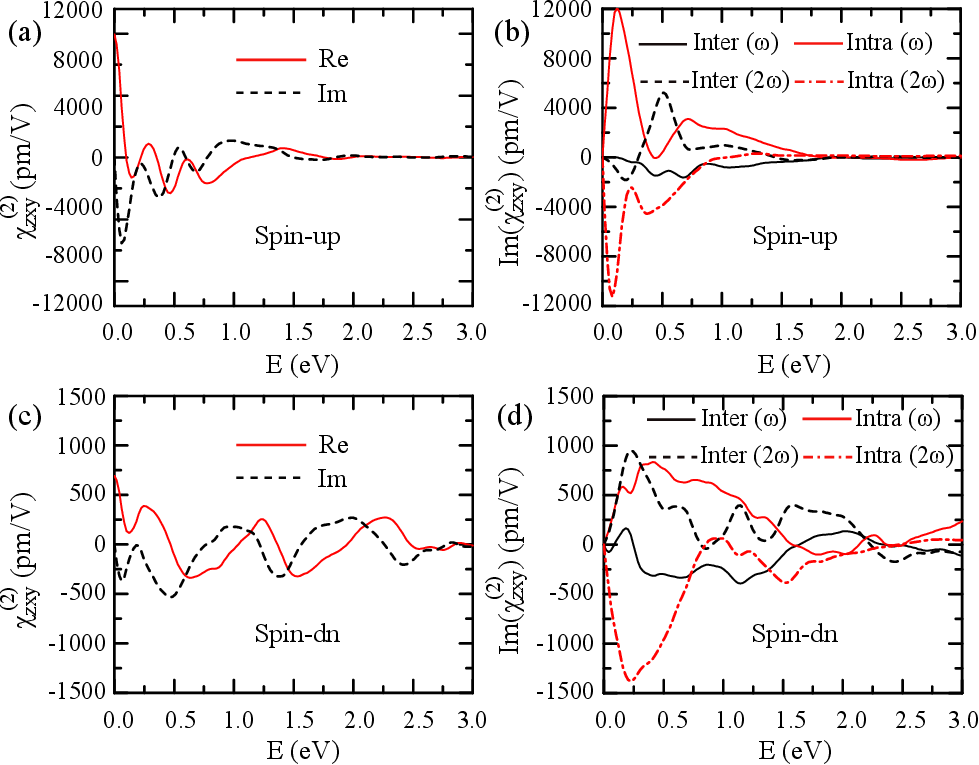}
\caption{(a) The real and imaginary parts of the SHG susceptibility $\chi_{zxy}^{(2)}$ from the spin-up channel in Mn$_{2}$RuGa without SOC.
(b) Calculated Im($\chi_{zxy}^{(2)}$) from inter($\omega$)/(2$\omega$) (black solid, black dashed curve)
and intra ($\omega$)/(2$\omega$) (red solid, red dashed-dotted curve) parts.
(c) and (d) are similar to (a) and (b), but from the spin-down channel.}
\label{fig7}
\end{figure}
Next, the absolute value of the SHG susceptibility $\chi_{zxy}^{(2)}$ for Mn$_{2}$RuGa without SOC is given in Fig. 6.
For the spin-up channel, the intensity at zero frequency is 9936.41 pm/V, which is smaller than that of $\chi_{xyz}^{(2)}$.
The secondary peak appears at 0.41 eV with an intensity of 3452.36 pm/V.
When the photon energy is larger than 0.59 eV,
the spectrum oscillates and then approaches to 0.
In contrast, the intensity of the spin-down $\chi_{zxy}^{(2)}$ is very small in the energy range from 0 to 3 eV.
Therefore, in the low energy region, the intensity of the spin-up $\chi_{zxy}^{(2)}$ is much stronger than that of the spin-down channel, which is similar to $\chi_{xyz}^{(2)}$.
This is because the spin-up channel has no band gap, while a gap in the spin-down channel limits SHG.

The real and imaginary parts of the spin-up channel for $\chi_{zxy}^{(2)}$ is shown in Fig. 7(a).
We decompose Im$\chi_{xyz}^{(2)}$ of the spin-up channel into the inter- and intraband contributions, as shown in Fig. 7(b).
It is found that the intraband contributions are dominant in the low energy region.
For the spin-down channel, the real and imaginary parts are very small, as shown in Fig. 7(c).
It mainly comes from the intraband and interband transitions, as shown in Fig. 7(d).
These results of both spin-up and spin-down channels are similar to $\chi_{xyz}^{(2)}$.

\subsection*{APPENDIX C: The SHG susceptibility for Mn$_{2}$RuGa with SOC}

\begin{figure}[bt]
\centering
\includegraphics[width=.45\textwidth]{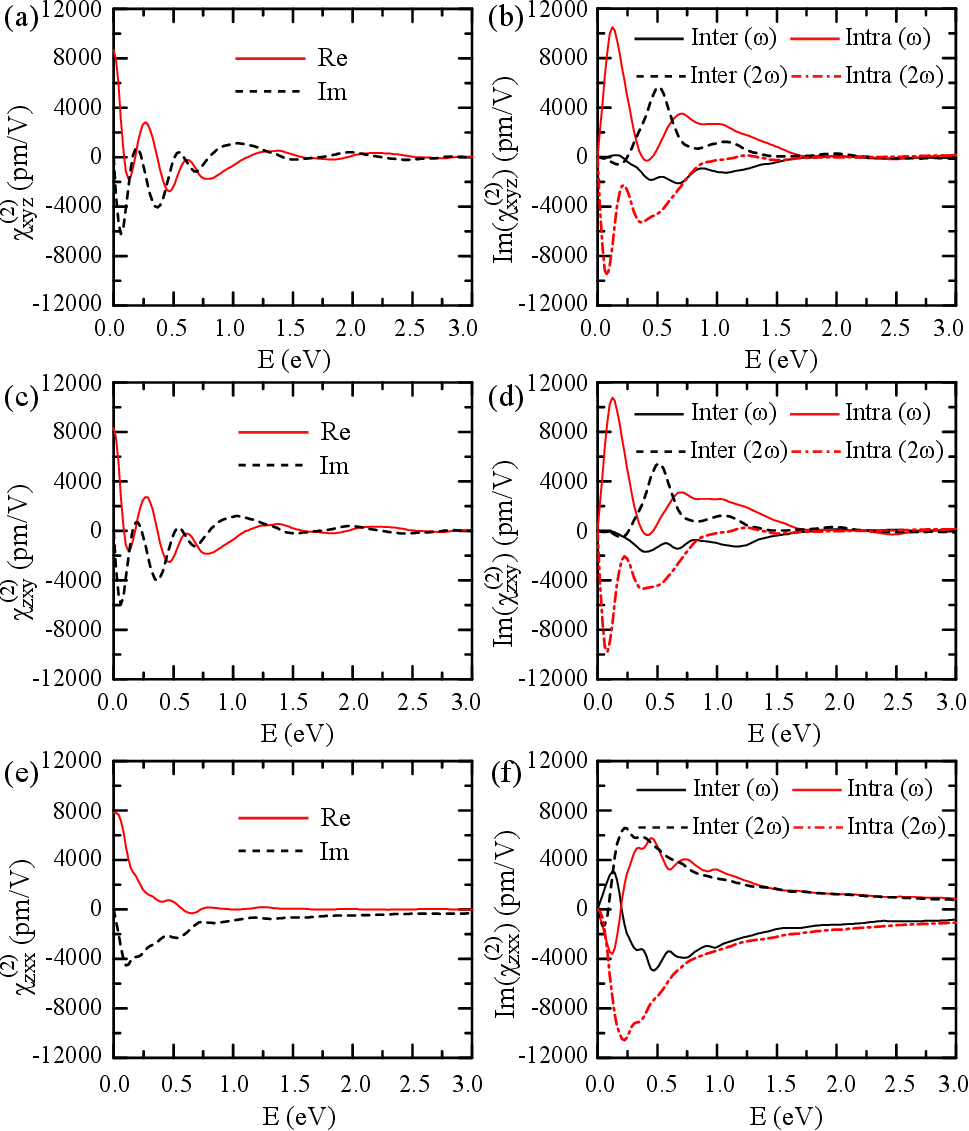}
\caption{(a) The real and imaginary parts of $\chi_{xyz}^{(2)}$.
(b) Calculated Im($\chi_{xyz}^{(2)}$) from inter($\omega$)/(2$\omega$) (black solid, black dashed curve)
and intra ($\omega$)/(2$\omega$) (red solid, red dashed-dotted curve) parts.
(c) and (d) are similar to (a) and (b), but for $\chi_{zxy}^{(2)}$.
(e) and (f) are similar to (a) and (b), but for $\chi_{zxx}^{(2)}$.}
\label{fig8}
\end{figure}
With SOC, $\chi_{xyz}^{(2)}$ and $\chi_{zxy}^{(2)}$ are similar
to the spin-up $\chi_{xyz}^{(2)}$ and $\chi_{zxy}^{(2)}$, respectively, as shown in Figs. 8(a)-8(d).
It shows that SOC has little effect on these two components.
On the contrary, the reduced symmetries induce $\chi_{xxz}^{(2)}$ and $\chi_{zxx}^{(2)}$,
showing different characteristics.
The real and imaginary parts of $\chi_{zxx}^{(2)}$ are shown in the red and black lines in Fig. 8(e).
It is contributed by inter- and intraband transitions, as shown in Fig. 8(f).
Compared with the spin-polarized $\chi_{xyz}^{(2)}$ and $\chi_{zxy}^{(2)}$, the interband contribution increases,
which is similar to $\chi_{xxz}^{(2)}$.

\clearpage

\end{document}